\documentclass[twocolumn,prd,showpacs
 amsmath,amssymb,
 aps, physrev,
]{revtex4-2}
\usepackage{graphicx}% Include figure files
\usepackage{caption}  % 标题格式控制
\usepackage{float}    % 图片位置控制
\usepackage{subcaption}
\usepackage{dcolumn}% Align table columns on decimal point
\usepackage{bm}% bold math
\usepackage{tabularx} % 用于调整表格宽度
\usepackage{booktabs} % 用于美化表格线条（可选）
\usepackage{amsmath}  % 用于数学符号
\usepackage{array} % 用于调整列宽
\usepackage{ragged2e}
\usepackage[utf8]{inputenc}
\usepackage{natbib}
\usepackage{ads}
\usepackage{xcolor} %

\begin{document}

\preprint{APS/123-QED}

%\title{\textbf{End to end simulation for axion-photon resonance conversion in magnetars} 
\title{Smoking gun signature from axion and the constraints with radio telescopes}

\author{Zixuan Liu}
\affiliation{Shanghai Astronomical Observatory, Chinese Academy of Sciences, Nandan Road 80, Shanghai 200030, China.}
\affiliation{School of Astronomy and Space Science, University of Chinese Academy of Sciences, No. 19A Yuquan Road, Beĳing 100049, China}

\author{Jiajun Zhang}%
\email{Contact author: jjzhang@shao.ac.cn}
\affiliation{Shanghai Astronomical Observatory, Chinese Academy of Sciences, Nandan Road 80, Shanghai 200030, China.}
\affiliation{Key Laboratory of Radio Astronomy and Technology (Chinese Academy of Sciences), A20 Datun Road, Chaoyang District, Beijing, 100101, China} 


\date{\today}% It is always \today, today,
          
\begin{abstract}
%As an elegant solution to the strong CP problem and a promising cold dark matter candidate, the laboratory detection of axions has long been limited by their extremely weak coupling. Leveraging the Primakoff effect of axions in magnetar magnetospheres, indirect astronomical searches for axion-photon coupling can be performed in the radio band. This study refines existing theories of resonant axion-photon conversion and constructs an end-to-end signal model linking dark matter spatial distribution to radio observables. Numerical simulations for QCD axions in the microelectronvolt ($\mu$eV) mass range show that under typical pulsar parameters, radio emissions generated via resonant conversion for axions with masses between 1-10$\mu$eV can be detected by China's Five-hundred-meter Aperture Spherical radio Telescope (FAST), reaching the coupling strength upper limits of QCD axion benchmark models. For sub-$\mu$eV ($<$1$\mu$eV) axions, future multi-band observations with the Square Kilometre Array (SKA) could achieve full coverage of the QCD theoretical prediction line. The end-to-end framework established here provides a transferable theoretical template for axion searches in the radio band, while quantitatively demonstrating the unique potential of next-generation facilities like FAST and SKA in probing fundamental particle physics.
Axions are an elegant solution to the strong CP problem for particle physics and a promising dark matter candidate. They can convert into photons under a strong magnetic field, while magnetars with extreme magnetic fields are natural labs for axion detection. Radio telescopes can detect the radio emission from axion-photon conversion near magnetars. In this study, we have refined the calculation of axion-photon conversion and developed the matched filtering integration method to largely improve the signal-to-noise ratio. We validate our method using end-to-end simulation and real observational data from TMRT. A new constraint is set with only 3596 seconds of observations with TMRT. Using 10 hours of observation with the high-frequency receiver in FAST or SKA, we can reach the theoretical coupling constant prediction for the axion mass range from 1$\mu$eV to 100$\mu$eV. We validate the possibility of axion detection with radio telescopes and avoid spectrum confusion.
\end{abstract}

\maketitle

%\tableofcontents

\section{\label{sec:level1}INTRODUCTION}

The nature of dark matter remains one of the most profound puzzles in modern physics\cite{2007JPSJ...76k1017P}. While weakly interacting massive particles (WIMPs) once stood as the leading candidate\cite{2025EPJC...85..490I}, their null detection has shifted attention to axions\cite{adams2023axiondarkmatter,sharma2024axionphysicsstringtheory}---a compelling solution rooted in quantum chromodynamics (QCD). The axion's necessity arises from the Peccei-Quinn mechanism\cite{1977PhRvL..38.1440P,1977PhRvD..16.1791P,Peccei:1996ax,Peccei:2006as,Kim:2008hd,PhysRevLett.40.279,PhysRevLett.40.223,Berezhiani:1989fp}, which elegantly resolves the CP-violation problem in QCD and predicts a particle with feeble but detectable couplings to photons\cite{PhysRevD.37.1237,PhysRevD.52.3132,2020PhR...870....1D}.

Axion-photon coupling is used to detect axions in ground laboratories. Numerous experiments employ resonant cavities in strong magnetic fields to detect axions, establishing progressively improved upper bounds on the axion-photon coupling strength $g_{a\gamma\gamma}$ across different axion mass ranges. Currently the ground axion search experiments detecting axion-photon coupling include  ADMX\cite{2025arXiv250407279A}, CAPP\cite{2024PhRvX..14c1023A}, HAYSTAC\cite{2025PhRvL.134o1006B}, RBF-UF\cite{PhysRevLett.59.839,PhysRevD.64.092003} and CAST\cite{2025arXiv250505909A}.

Astrophysical environments, particularly magnetars with their extreme magnetic fields ($B \sim 10^{13-15}\,\text{G}$), offer a natural laboratory for axion detection. Here, axion-photon resonance conversion occurs when the magnetar's plasma frequency matches the axion mass, significantly enhancing the probability of axion-photon conversion\cite{Pshirkov:2007st,Masaki_2017,Millar_2021,PhysRevD.110.015018}. This process mirrors laboratory cavity experiments but leverages natural magnetic fields to probe a broader axion mass range (1--100\,$\mu$eV)\cite{1978JETPL..27..502V}. Currently, advanced models employing Monte Carlo-based ray-tracing simulations have been developed to model the complex behavior of axion-converted photons within magnetospheres, revealing distinctive spectral, spatial, and temporal characteristics of the emission\cite{2021PhRvD.104j3030W, 2024PhRvD.109b3015T}.

The photon signals from astrophysical axion conversion naturally fall within radio telescope observational frequencies, making them theoretically detectable. People have raised forecasts for the detection\cite{PhysRevD.52.3132,Pshirkov:2007st,Masaki_2017,Millar_2021,PhysRevD.110.015018,2018PhRvD..97l3001H}. Recently Meerkat put a constraint using the observation of PSR J2144-3933\cite{PhysRevD.108.063001}. However, radio telescopes face both confusion limits (from source crowding) and sensitivity limits.  In this paper, we discovered a smoking gun signature from axion, which can help overcome the confusion limit. If provided achievable observing time of radio telescopes like FAST or SKA, we will be able to reach the prediction by KSVZ model\cite{1979PhRvL..43..103K,SHIFMAN1980493} and DFSZ model\cite{Zhitnitsky:1980tq,DINE1981199,PhysRevD.107.095020}. 

%This work establishes an end-to-end template for axion-photon conversion in magnetar magnetospheres, simulating the expected radio signals featuring temporal modulation and narrowband spectral characteristics. Our work bridges theoretical predictions with radio telescope observables and validates the feasibility of practical search methods. By incorporating parameters from different radio telescopes, we provide projections for observational prospects and potential constraints on the axion-photon coupling constant $g_{a\gamma\gamma}$. This study offers a comprehensive pathway toward breakthroughs in the QCD axion parameter space. 

\section{Axion-photon coupling}
%\subsection{Coupling of Axions and Photons}
The axion field couples to the electromagnetic field\cite{2020PhR...870....1D}.The axion-photon coupling term can be expressed as
\begin{equation}
L_{a\gamma\gamma}=\frac{1}{4}g_{a\gamma\gamma}aF_{\mu \nu} \tilde{F}^{\mu \nu}=-g_{a\gamma\gamma}a\textbf{E}\cdot\textbf{B},
\label{eq:lagrange}
\end{equation}
where $g_{a\gamma\gamma}$ represents the axion-photon coupling strength. Axions can decay into two photons via a triangle loop diagram\cite{2020PhR...870....1D}. When one photon leg is treated as an external electromagnetic field, the Feynman diagram describes axion-to-photon conversion in an external EM field.
%\subsection{axion-photon resonance conversion}

Applying perturbation theory for coupled eigenstates, the axion-photon conversion probability in an external magnetic field is derived as\cite{2018PhRvD..97l3001H}
\begin{equation}
    P_{a \rightarrow \gamma}(z)=\sin ^2(2 \tilde{\theta}(z)) \sin ^2\left(\frac{\Delta k z}{2}\right),
\label{eq:probability}
\end{equation}
where $\tilde{\theta}$ is the mixing angle and $\Delta k$ is the wave vector difference between the two new eigenstates after Bogoliubov transformation. The mixed eigenstates propagate along the z-axis.

The calculation assumes a time-independent dipole field B(r)\cite{PhysRevD.37.1237}. When the photon's effective mass in the external EM field matches the axion mass, the system's two eigenstates become highly degenerate, leading to maximal mixing. The system achieves resonant axion-photon conversion, significantly enhanced the conversion probability\cite{PhysRevD.110.015018}.
\section{Resonance conversion in magnetosphere}
This study accounts for the periodic dipole magnetic field of magnetars. For a given point in the magnetosphere along the line of sight, the magnetic field strength varies periodically with time, consequently modulating the axion-photon conversion probability and inducing observable variations in photon flux. This should manifest in our observations as axion signals exhibiting periodic temporal characteristics similar to those of the magnetar. However, directly incorporating time-dependent dipole fields into resonance modeling is computationally complex. Furthermore, the inability to precisely determine the inclination angle $\alpha$ prevents accurate localization of the anisotropic effects caused by magnetic field periodicity\cite{2022PhRvD.105b1305B,2012hpa..book.....L,Gil1984AA,2023MNRAS.520.4801J,2019MNRAS.485..640J,2015MNRAS.446.3367R}. To simplify calculations, the template employs a static dipole field\cite{2018PhRvD..97l3001H}, while applying periodic temporal modulation during subsequent signal simulations to approximate both the periodicity and anisotropy. 
It should be noted that realistic magnetar fields likely possess more complex structures\cite{2018PhyU...61..353B}, including pulsar current sheets and the twisted magnetic configurations characteristic of strongly magnetized neutron stars. Recent studies have compared the unique properties of the converted signals under two different magnetosphere models. Their simulations show that the spatial distribution of the converted emission exhibits significant anisotropy and may possess time-evolving features\cite{2025arXiv250520450R}. Given the periodic rotation of the magnetar itself, we consider it reasonable to assume that the observed signal should exhibit temporal periodicity. Within limited observational time windows, this periodicity is unlikely to be significantly altered by potential temporal evolution of the magnetosphere. These more sophisticated scenarios remain subjects for future fine template matching.

%\subsection{Photon plasma effect and QED vacuum polarization}
In the strong magnetic field of a magnetar, the vacuum exhibits polarization effects\cite{PhysRevD.37.1237,1971AnPhy..67..599A}. For the magnetic field strengths below $B_{\text {crit}}=4.4 \times 10^{13} \mathrm{G}$, QED effects can be neglected\cite{2018PhRvD..97l3001H,PhysRevD.37.1237,PhysRevD.37.2039}, and the Goldreich-Julian (GJ) model remains applicable\cite{1969BAAS....1R.242G}. When photons propagate through the plasma, collective oscillations of electrons under the electric field occur, with their characteristic frequency known as the plasma frequency $\omega_p$. We have considered the photon dispersion relation in plasma under the combined effects of gravity and strong magnetic fields\cite{2001PhRvE..64b7401G,2019IJMPD..2840013T}. By taking the non-relativistic limit and assuming a pure electron-positron plasma\cite{PhysRevD.108.063001}, we derive the simplified plasma frequency that represents the photon's effective mass. Then, resonance conversion between axions and photons occurs when the photon's effective mass equals the axion mass ($m_\gamma^2=m_a^2$). We numerically solve the equation $m_\gamma^2=m_a^2$ to determine the resonance radius $r_{\rm res}$.

For magnetars with surface fields exceeding $10^{13}\,\text{G}$, vacuum
polarization effects remain relatively weak, %($Q_{\rm QED} \sim 0.0036\omega^2$ at $B_0 \sim 10^{14}\,\text{G}$)
while Landau quantization effects become dominant\cite{1930ZPhy...64..629L}. The intense magnetic field
confines electron motion to the lowest Landau level, enhancing local
electron densities by factors of 2--10. When accounting for additional
photon pair cascade processes, this density enhancement can increase by
3--5 orders of magnitude, significantly altering the plasma
characteristics in the magnetar magnetosphere.
%\subsection{axion-photon resonance conversion region}

\begin{figure*}[htbp]
    \centering
    \includegraphics[width=\textwidth]{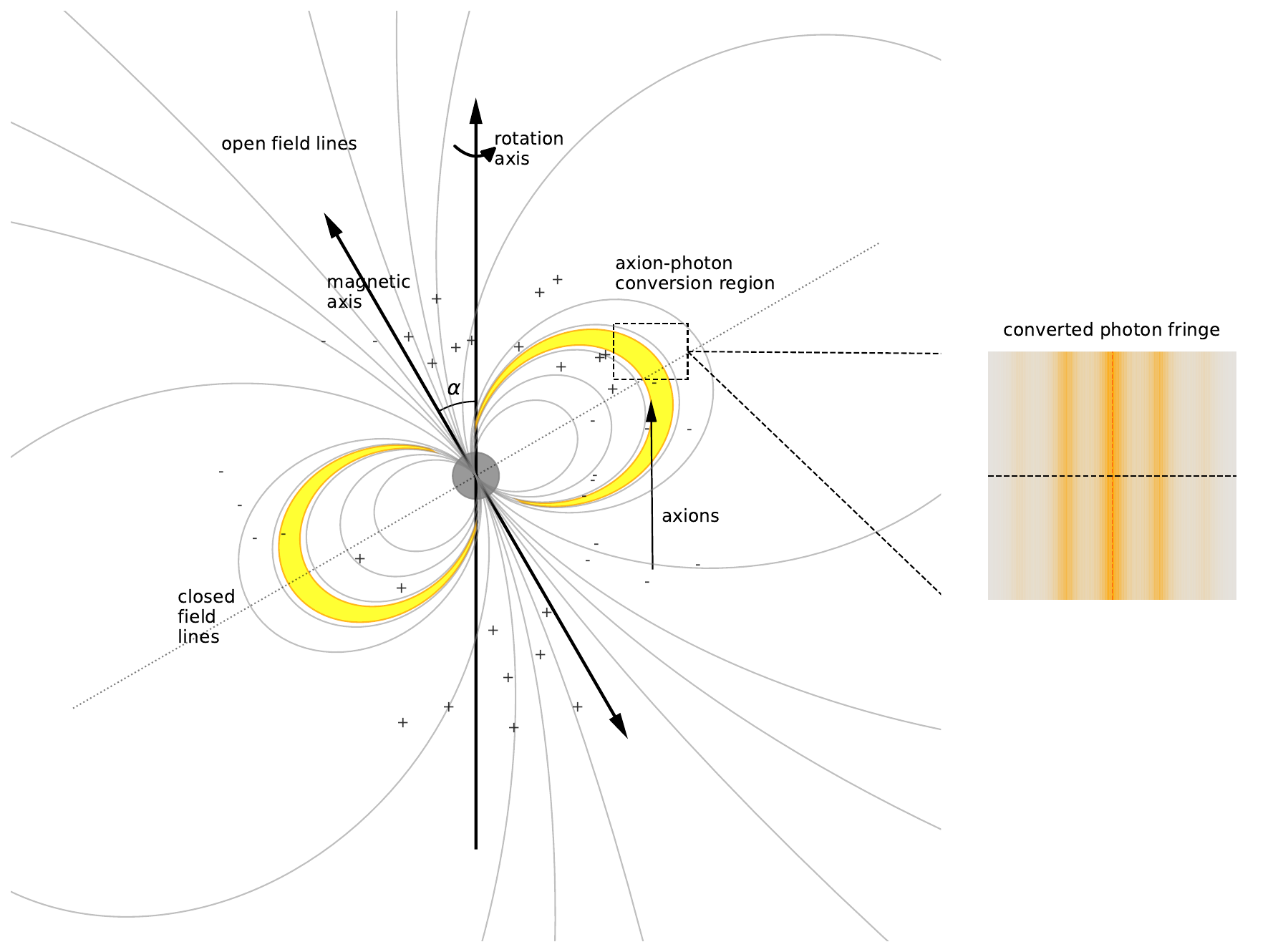}  % 填满版心宽度
    \captionsetup{justification=raggedright,singlelinecheck=false}
    \caption{Schematic diagram of axion-photon resonance conversion in magnetars' magnetosphere. The yellow-shaded zone in the schematic represents the axion-photon resonant conversion region, where magnified views reveal characteristic fringes from the axion-photon oscillations.}  % 图注
    \label{fig:schema}   % 标签用于引用
\end{figure*}

To ensure the adiabaticity of axion-photon resonant conversion, we define the resonance region near the resonance radius $r_{\rm res}$ following \citet{2018PhRvD..97l3001H}. In conversion region, we set the origin of the $z$-axis at $r_{\rm res}$ with $z = r - r_{\rm res}$. As described in Eq.~\ref{eq:probability}, the first term represents the external low-frequency envelope that determines the overall conversion probability from axions to photons, while the second term corresponds to the internal high-frequency oscillation that governs whether complete oscillation between axions and photons can occur. Within the resonance region, the total conversion probability evolves gradually along the external envelope as $z$ varies. By dividing the resonance region into thin layers, each containing one high-frequency oscillation cycle, we observe partial axion-to-photon and photon-to-axion conversions within each layer. The net converted photon flux is obtained by integrating these contributions along the $z$-axis across the entire effective resonance region. In Fig.~\ref{fig:schema}, we schematically illustrate the axion-photon resonant conversion process within a magnetar's magnetosphere. Theoretically, with sufficiently high resolution, one could observe distinctive fringes of converted photons within the resonance conversion region.

\section{axion signal simulation}
We developed a refined template to calculate axion-photon conversion in magnetars. This template does not adopt the highly complex magnetosphere models presented in the latest studies\cite{2025arXiv250520450R}, as signals produced by such detailed models are strongly model-dependent—and the true magnetospheric structure of magnetars remains uncertain. Instead, we aim to employ a more model-independent approach in the initial signal search stage. We argue that the temporal periodicity of the observed signal, induced by magnetar rotation, satisfies this requirement. More sophisticated models would be applied at later stages for signal confirmation and detailed interpretation. Therefore the template accounts for the time-dependent periodicity of axion-photon conversion. Given a set of magnetar parameters, it calculates the converted photon flux and generates end-to-end simulated signals. This framework serves two primary purposes: (1) validating actual axion search methodologies, and (2) providing precise templates for subsequent matched-filter analysis.

We derived the template of the simulation axion-to-photon signal function:
\begin{equation}
\begin{aligned}
& S_\gamma(t, \nu) = A \cdot \frac{\Gamma / 2}{(\nu-\nu_0)^2 + (\Gamma / 2)^2} \\
& \cdot \exp \left[-\frac{(t \bmod P - t_0 - 4.15 \times 10^6 \cdot \mathrm{DM} / \nu^2)^2}{2 \sigma_t^2}\right],
\label{eq:template}
\end{aligned}
\end{equation}
where the amplitude $A=\frac{d E / d t}{4 \pi d^2 \Delta \nu}$ corresponds to the calculated photon conversion flux. The expression $t \bmod P$ represents the modulo operation of time t with respect to period P. The spectrum is modeled using a Lorentzian function, with the line width $\Delta\nu=\nu_{peak}v_{dis}$. To capture the temporal periodicity of the magnetar's magnetic field, the temporal profile is fitted with a Gaussian periodic function, incorporating time delay effects from dispersion measure (DM). The DM along the photon propagation path includes contributions from both the magnetospheric plasma and the interstellar medium. The signal function's temporal periodicity inherently reflects the anisotropic nature of the magnetospheric magnetic field.
\section{Verification of search Method}
For pulsar J1022+1001, using parameters from the ATNF Pulsar Catalogue\cite{2005AJ....129.1993M}, our template calculates a total converted photon flux of 1.371 $\mu$Jy for axions with a mass of 10 $\mu$eV and a photon coupling constant of order $10^{-13} GeV^{-1}$. We verify the axion search methodology using TMRT's S-band (2200-2300 MHz) observations of pulsar J1022+1001. We injected simulated axion-photon conversion signals with a predicted flux of 1.371 $\mu$Jy, which remained undetectable in the original 481-second integrated data. Purely considering thermal noise and according to the sensitivity of the Tianma telescope\cite{2024AstTI...1..239L}, the signal needs to be artificially amplified by 1000 times to reach detectable levels. This would require impractical integration times exceeding O($10^3$) hours, making it practically unachievable for actual observations.
%\subsection{confusion limit solution}
However, in addition to thermal noise, radio telescopes are also subject to the confusion limit. Within the telescope's beam, other astrophysical sources contribute to the observed spectrum, creating an unresolved background. To overcome this confusion limit, we utilize the temporal periodicity of the axion signal and employ time-weighted integration methods. Our method can mitigate both stable interference sources and standing waves. Given the magnetar's period, we apply a weighted template across each cycle: assigning a weight of (1–t) to the expected signal window of width t and –t to the remaining phase bins. This suppresses non-periodic noise while enhancing coherent signals at their correct phase and width. By scanning different window widths (t) and phase offsets, we identify the parameter combination yielding the strongest response.

\begin{figure}[htp]
    \centering
    \includegraphics[width=0.45\textwidth]{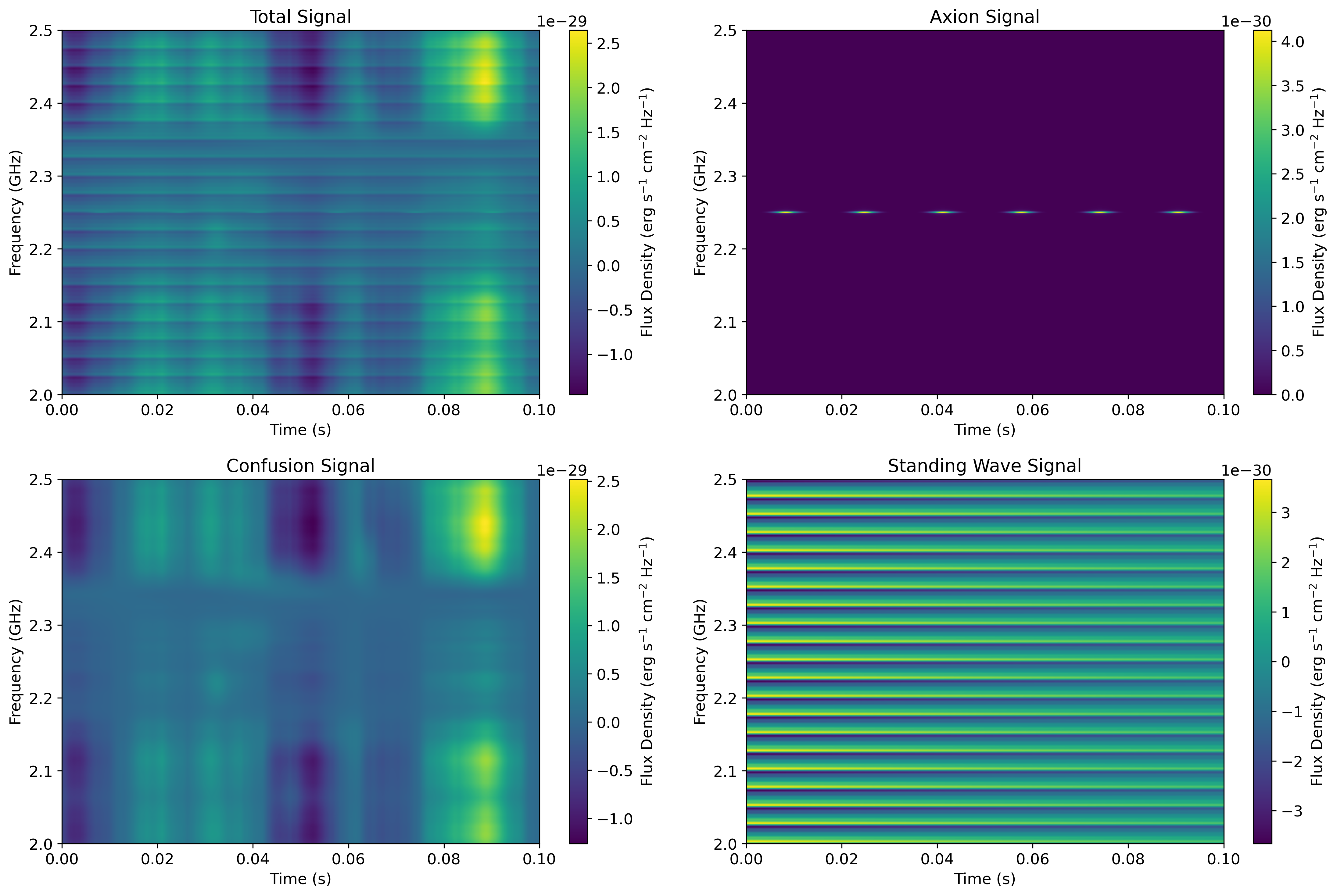}
    \centering
    \includegraphics[width=0.45\textwidth]{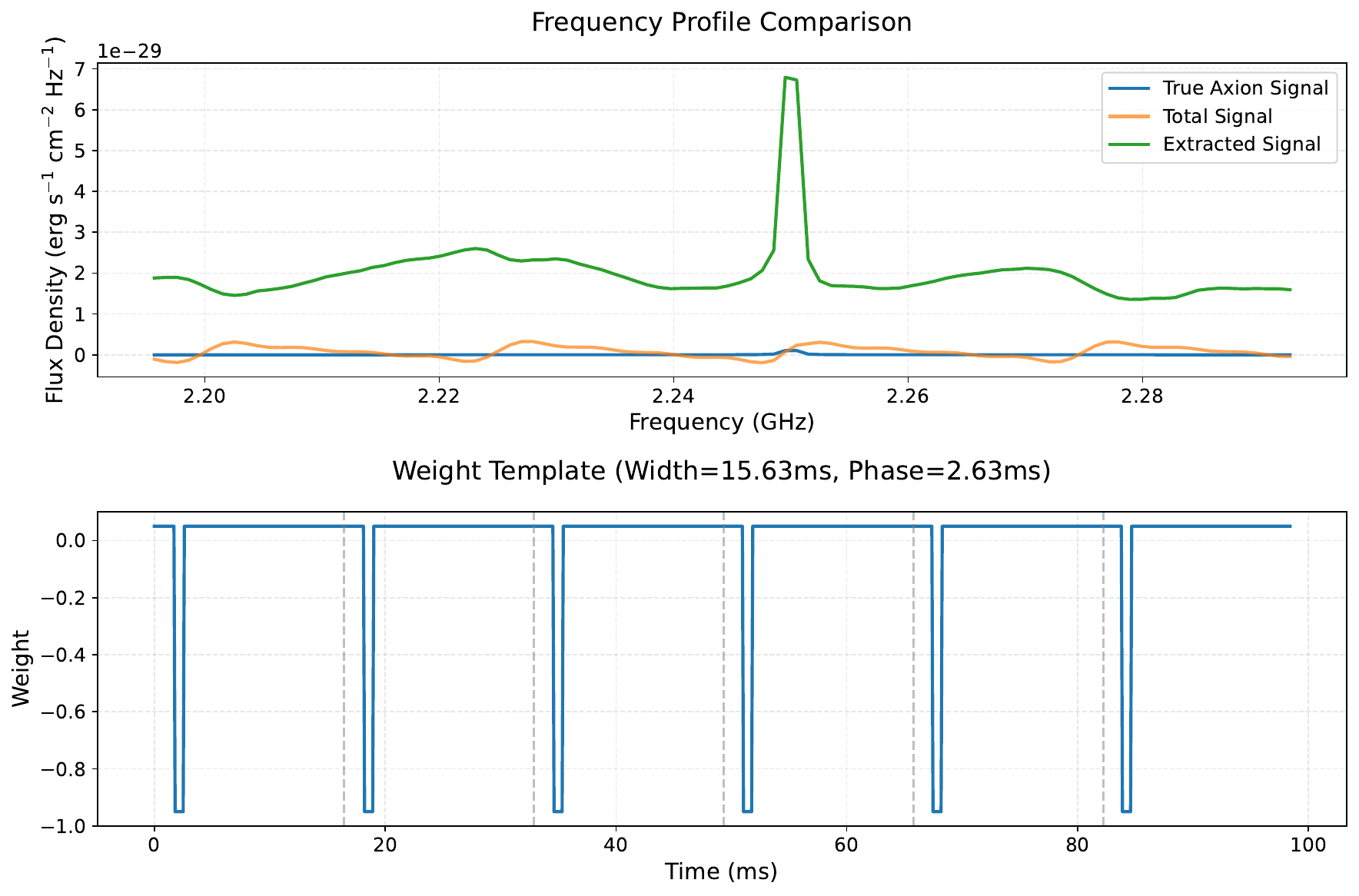}
    \captionsetup{justification=raggedright,singlelinecheck=false}
    \caption{The upper panel shows the simulated signal in the beam of the radio telescope, with the simulated overall signals in the upper left corner, the simulated periodic axion signal in the upper right corner, the simulated confusion signal in the lower left corner and the simulated telescope standing wave in the lower right corner. The lower panel is the comparison between the simulated axion signal extracted by time-weighted integration and the total signal. This method can effectively extract the known periodic signal from the interference sources. The bottom panel shows the weight template used.}
    \label{fig:confusion-solution}
\end{figure}

We show the simulations of radio telescope signals and the results of periodic signal extraction in Fig.~\ref{fig:confusion-solution}. The upper panel displays simulated radio signals: the composite signal shown in the upper-left represents the combined signals from the other three subplots; the upper-right shows the simulated axion signal; the lower-left illustrates simulated confusion signals; and the lower-right depicts simulated telescope standing waves. The time-weighted integration method can directly extract periodic axion signals from this complex combination. The lower panel demonstrates the extraction results, the upper subpanel compares the effect of weighted integration and usual time-domain integration, while the lower subpanel shows the corresponding weighting template. These results demonstrate that for periodic signal searches with predefined periods, our weighted integration method significantly enhanced the effectiveness of axion-signal detection.
%\subsection{signal function matched filter}

When the signal template already encodes the expected temporal and spectral features,(e.g., Eq.~\ref{eq:template}), matched filter can precisely identify axion-to-photon signals while providing optimal signal-to-noise ratio SNR improvement\cite{PhysRevD.108.063001}. However, this approach is highly sensitive to the model parameters, risking signal mis-detection. Our time-weighted integration approach enables model-independent axion searches with higher computational efficiency. For practical detection, we propose a two-stage strategy: applying time-weighted integration to identify candidate signals, followed by template matching to perform model parameter inference.

Beyond verifying its ability to suppress confusion, we further evaluated the quantitative advantage of the time-weighted integration method. By comparing the signal-to-noise ratio (SNR) obtained through direct integration and time-weighted integration under identical input flux conditions (for both signals and thermal noise), we found that the SNR achieved with the time-weighted method was approximately twice as large. This enhanced SNR performance indicates that, in practical implementations, it is more appropriate to use SNR thresholds—rather than the telescope sensitivity formula—to establish the upper limits of detectable flux. This approach better reflects the true extraction efficiency of the signal processing method.
\section{Results}
Theoretically, axion-to-photon signals would be detectable if the signal flux exceeds the radio telescope's sensitivity threshold. The sensitivity of a radio telescope, representing the minimum detectable flux density, is expressed as
\begin{equation}
S_{\rm min} = \frac{2k_BT_{\rm sys}}{A_{\rm eff}\sqrt{\Delta B\Delta t}}
\end{equation}
where $A_{\rm eff} = \eta(\pi D^2/4)$ denotes the effective collecting area with D being the antenna diameter and $\eta$ the aperture efficiency, $\Delta t$ is the integration time, $T_{\rm sys}$ corresponds to the system temperature across the entire receiver bandwidth, and $\Delta$B is the total bandwidth, with the calculated $S_{\rm min}$ representing the average sensitivity over the full receiver bandwidth.
%\subsection{Focus}

When no axion signal is detected in the data, the template can derive upper bounds on the axion-photon coupling constant by incorporating the radio telescope's sensitivity parameters. For the TMRT telescope, its S-band(2200-2300MHz) which has $T_{\rm sys}=33K$ with an observation time of 10 hours yields $S_{\rm min}=7.62 \times 10^2 \mu Jy$ \cite{2024AstTI...1..239L}, corresponding to an axion mass range of 9.1-9.5$\mu$eV and giving an upper bound $g_{a\gamma\gamma}<10^{-9}GeV^{-1}$; for the Q-band(35-50GHz) receiver with $T_{\rm sys}=70K$ \cite{2024AstTI...1..239L}, the calculated $S_{\rm min}=5.01\mu Jy$ corresponds to an axion mass range of 144.75-206.78$\mu$eV and gives an upper bound $g_{a\gamma\gamma}<10^{-14}GeV^{-1}$. If FAST is equipped with a high frequency receiver, considering FAST's collecting area with a system temperature $T_{\rm sys} = 20$\,K, a receiver bandwidth of 1GHz and an observation time of 10 hours, we obtained a minimum detectable flux density $S_{\rm min} = 6.69 \times 10^{-2}\,\mu$Jy in the target frequency range, which constrains the axion-photon coupling to $g_{a\gamma\gamma} < 1.14 \times 10^{-13}$\,GeV$^{-1}$ for 10\,$\mu$eV axions and $g_{a\gamma\gamma} < 3.51 \times 10^{-13}$\,GeV$^{-1}$ for 100\,$\mu$eV axions. For SKA-mid band 5b observed for 10 hours\cite{2019arXiv191212699B}, theoretical calculations yield a more stringent upper bound on the axion-photon coupling about $g_{a\gamma\gamma}<10^{-14}GeV^{-1}$.
%\section{data search}

We performed a periodic signal search on 3595.94209667 s of observational data from Pulsar J1809-1943 using validated methods. The data were obtained from the Tianma Telescope's S-band (2200.20-2299.80 MHz) and X-band (8201.56-8996.48 MHz) observations\cite{2024AstTI...1..239L}, with pulsar parameters provided by The ATNF Pulsar Catalogue\cite{2005AJ....129.1993M}. The temporal periodicity filter scanned 649 periods of data but detected no significant signals. The filters yielded signal-to-noise ratios of 3.3 (S-band) and 1.2 (X-band), which were dominated by strong, unstable RFI (radio frequency interference) rather than genuine axion signatures, and thus do not indicate potential axion signals.

Based on the Tianma Telescope's sensitivity parameters, and adopting a conservative signal-to-noise threshold of $5\sigma$, we evaluated the upper bounds for the axion-photon coupling. Compared to using the sensitivity formula, the $5\sigma$ signal-to-noise threshold yields more conservative constraints. Simultaneously, by employing time-weighted integration, which reduces the impact of noise and the required input flux by a factor of 4–5, the coupling constraint can be correspondingly tightened. Taking all these improvements into account, the overall coupling constant upper bounds are given at $g_{a\gamma\gamma} < 8.139\times10^{-10} GeV^{-1}$ and $g_{a\gamma\gamma} < 3.045\times10^{-10} GeV^{-1}$ for the respective bands.

\section{conclusion}
In this paper, we found that the axion-photon resonant conversion effect is suitable for detecting axions around magnetars using a radio telescope. In the magnetosphere of a magnetar, our calculation suggested that there are axion-photon conversion fringes. This is a smoking gun signature of axion-photon resonant conversion, but far beyond our detection capability. Besides the spatial signature, we also discussed that due to the periodicity of magnetic fields around a magnetar, the signal from axion-photon conversion can be modulated by the periodicity. This temporal signature is also a smoking gun for axion detection. We provide the computational framework and the signal template.

Using the template, we generate simulated photon signals, which we have used to validate the feasibility of searching for axion-photon conversion signatures. We proposed a two-stage approach: applying periodic weighted integration to extract candidate signals matching the magnetar's periodicity, followed by matched filtering using precisely generated signal templates. Our simulation studies reveal that our weighted integration method is effective against the confusion limit. Our methodology enables detection of axion signals up to the telescope's sensitivity limit and provides robust upper limits in cases of non-detection.

For axion masses of 10 $\mu$eV and 100 $\mu$eV, our forecast shows that the framework achieves theoretical sensitivity limits of $g_{a\gamma\gamma} < 1.14\times10^{-13}$ GeV$^{-1}$ and $g_{a\gamma\gamma} < 3.51\times10^{-13}$ GeV$^{-1}$ respectively. We subsequently applied our periodic weighted integration technique to preliminary searches of Pulsar J1809-1943 in both S-band (9.1-9.5 $\mu$eV) and X-band (33.9-37.2 $\mu$eV) observations, yielding null results with upper limits of $g < 8.139\times10^{-10}$ GeV$^{-1}$ and $g < 3.045\times10^{-10}$ GeV$^{-1}$ respectively. These constraints, while currently limited by instrumental sensitivity, demonstrate the framework's capability to detect axions.

\begin{figure*} % H参数强制图片位置
\centering
\includegraphics[width=0.95\textwidth]{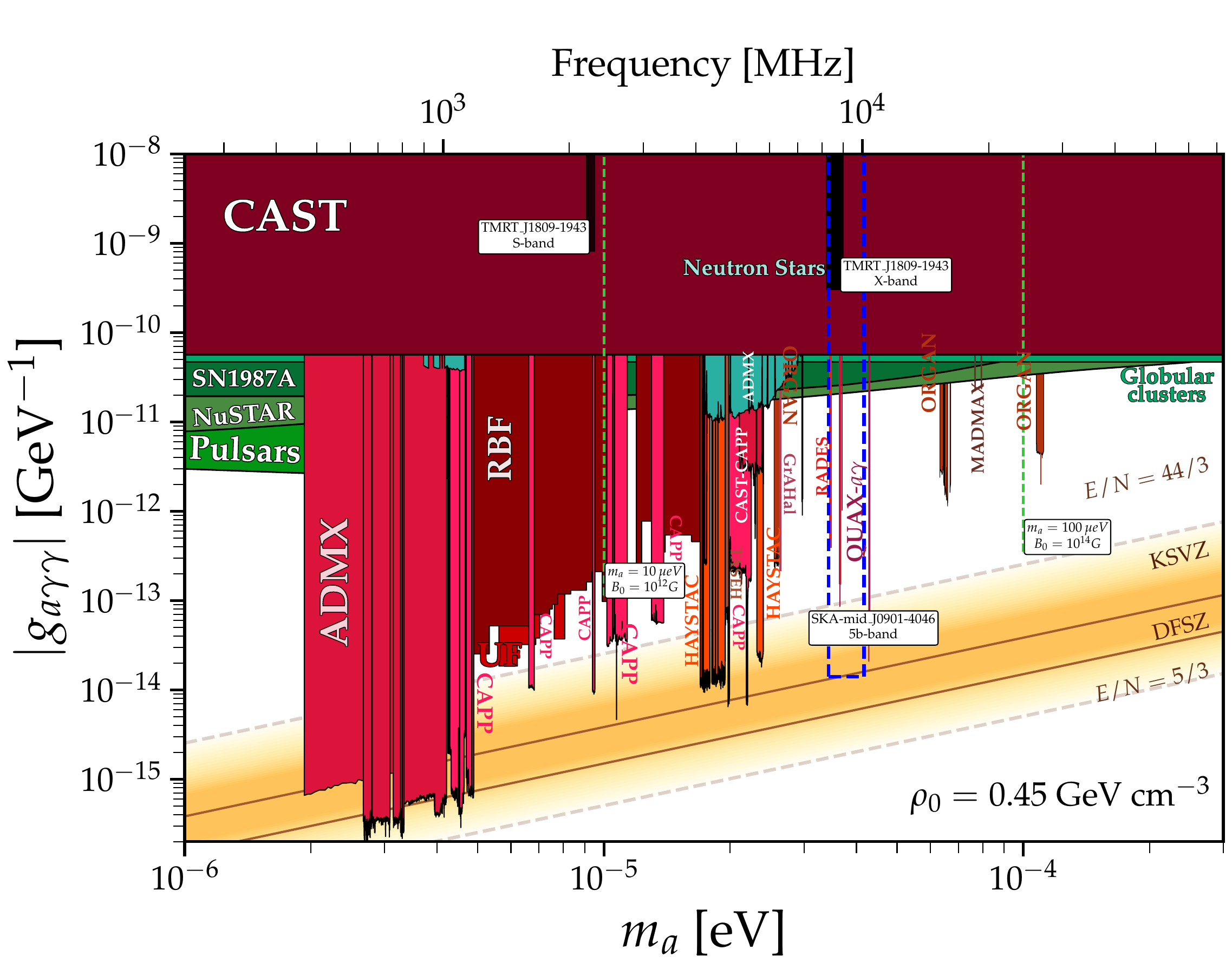}
\captionsetup{justification=raggedright,singlelinecheck=false}
\caption[Axion-photon coupling limits]{
Constraints on axion-photon coupling $g_{a\gamma}$ from various observation approaches\footnote{https://
cajohare.github.io/AxionLimits/}, including both forecasts and actual observational constraints from this work. The dashed line is the forecast constraints. The green dashed lines show the upper bound of the axion-photon coupling for axion masses of 10 $\mu$eV and 100 $\mu$eV, calculated using FAST's parameters as an example. The blue dashed line shows the upper bound calculated for SKA-mid's band 5b, using J0901-4046 as the observation source. The solid black region represents the coupling upper limits derived from the observational data search of J1809-1943 using TMRT, with the constraints provided by the S-band and X-band, respectively.
}
\label{AxionLimits}
\end{figure*}
\section{Acknowledgements}
This work is supported by Key Laboratory of Radio Astronomy and Technology (Chinese Academy of Sciences). We thank Yajun Wu, Zhen Yan, and Zhiqiang Shen for providing testing data from TMRT. JJZ acknowledges the startup funding provided by Shanghai Astronomical Observatory, CAS.
\appendix

\bibliography{axion}% Produces the bibliography via BibTeX.

\end{document}